\documentclass[onecolumn,10pt]{article}
\usepackage{amssymb}
\usepackage{times}

\newcommand{\hil}[1]{\mbox{$\mathcal{#1}$}}

\newcommand{\ket}[1]{| #1 \rangle}

\begin{document}
\date{}
\title{\textbf{Quantum Computation \\ and Pseudo-telepathic Games}}
   \vspace{1in}
   
   \author{ Jeffrey Bub\footnote{jbub@umd.edu}\\
\footnotesize Department of Philosophy, University of Maryland,
College Park, MD 20742}
\maketitle

\begin{abstract}
A quantum algorithm succeeds not because the superposition principle allows `the computation of all values of a function at once' via `quantum parallelism,' but rather because the structure of a quantum state space allows new sorts of correlations associated with entanglement, with new possibilities for information-processing transformations between correlations,  that are not possible in a classical state space.  I illustrate this with an elementary example of a problem for which a quantum algorithm is more efficient than any classical algorithm. I also introduce the notion of  `pseudo-telepathic' games and show how the difference between classical and quantum correlations plays a similar role here for games that can be won by quantum players exploiting entanglement, but not by classical players whose only allowed common resource consists of shared strings of random numbers (common causes of the players' correlated responses in a game).

\end{abstract} 
\bigskip

\section{Introduction}  
The power of quantum computation is commonly thought to derive from the superposition principle, which, in some sense, `allows the computation of all values of a function at once' via `quantum parallelism.' The claim is that a quantum computation is something like a massively parallel classical computation, for all possible values of a function. This appears to be Deutsch's view \cite{DeutschBook}: in an Everettian many-worlds interpretation of quantum mechanics,  the parallel computations can be regarded as taking place in parallel  universes. (For a critique, see \cite{Steane03}.) 

The basic idea can be put simply as follows: A quantum computer stores input values of a function in an input register and output values in an output register. In the simplest case of a Boolean function, $f:\{0,1\} \rightarrow \{0,1\}$, the input register is a qubit, a quantum system whose state can be represented by a vector in a 2-dimensional Hilbert space. Each of the  two possible input values for the function can be stored as one of two orthogonal quantum states in a standard basis called the computational basis, $\ket{0}_{i}$ or $\ket{1}_{i}$. The output register is also a qubit, and each of the two possible outputs can be stored  in one of two orthogonal states, $\ket{0}_{o}$ or $\ket{1}_{o}$. During the implementation of a quantum algorithm, the input and output registers can separately be in linear superpositions of the computational basis states, and the two registers can be in an entangled state.

If the input register is in a linear superposition of the two possible input states, and the output register is in the state $\ket{0}$:
\begin{equation}
(\frac{1}{\sqrt{2}}(\ket{0} + \ket{1})\ket{0}\footnote{The subscripts are dropped in the following. The order of the two states represents input and output.}
\end{equation}
a unitary transformation $\ket{0}\ket{0} \stackrel{U_{f}}{\rightarrow} \ket{0}\ket{f(0)}, \ket{1}\ket{0} \stackrel{U_{f}}{\rightarrow} \ket{1}\ket{f(1)}$ implementing the function $f$\footnote{To complete the definition of a unitary transformation, we would have to add: $\ket{0}\ket{1} \stackrel{U_{f}}{\rightarrow} \ket{0}\ket{1 \oplus f(0)}, \ket{1}\ket{1} \stackrel{U_{f}}{\rightarrow} \ket{1}\ket{1 \oplus f(1)}$, where $\oplus$ is Boolean addition.}  will induce the transformation:
\begin{equation}
\frac{1}{\sqrt{2}}(\ket{0}\ket{f(0)} + \ket{1}\ket{f{1}}) \label{eqn:parallelism}
\end{equation}
correlating the two possible input values with the corresponding output values in the---now entangled---state of the two registers.

Of course, while all input-output value pairs for the function $f$ are represented in the entangled state, these values are not accessible without a measurement, and a measurement in the computational basis $\ket{0}, \ket{1}$ will return just one of the input-output pairs, either $\{0,f(0)\}$ or $\{1,f(1)\}$, randomly. 

So, how does a quantum computation achieve a gain in efficiency relative to a classical computation? Essentially, a disjunctive (or global) property of a function, such as the period of a periodic function, is represented as a subspace in Hilbert space, which can then be efficiently distinguished from alternative disjunctive properties by a measurement (or sequence of measurements) that identifies the target disjunctive property as the property represented by the subspace containing the final state produced by the algorithm. While `quantum parallelism' might be involved as a stage in a quantum algorithm, rather than `computing all values of a function at once,' a quantum algorithm works by generally avoiding the computation of \emph{any} values of the function at all. The point of the procedure is precisely to \emph{avoid the evaluation of the function} in the determination of the disjunctive property, in the sense of producing output values for given input values of the function. The fact that alternative disjunctive properties can be efficiently distinguished in this way, without requiring the evaluation of the function for individual values of the input, via a quantum algorithm but not via a classical algorithm, depends on structural differences between quantum and classical correlations associated with entanglement in the quantum case and common cause correlations in the classical case, and on  the possible evolutions between correlations associated with information-processing in a computation, and it is these differences that are exploited to achieve the relative efficiency of the quantum algorithm.

I illustrate this with Deutsch's XOR problem in section~\ref{XOR}, an elementary example of a problem for which a quantum algorithm is more efficient than any classical algorithm. In section~\ref{pseudo}, I introduce the notion of  `pseudo-telepathic' games and show how the difference between classical and quantum correlations plays a similar role here for games that can be won by quantum players exploiting entanglement, but not by classical players whose only allowed common resource consists of shared strings of random numbers (common causes of the players' correlated responses in a game).
\section{A Quantum Computation}
\label{XOR}

In Deutsch's XOR problem \cite{Deutsch1985}, a `black box' or oracle  computes a Boolean function $f:  
\{0,1\} \rightarrow  \{0,1\}$. The problem is to determine whether the function is `constant' (takes the same value for both inputs) or `balanced' (takes a different value for each input). The properties `constant' and `balanced' are two alternative disjunctive properties of the function $f$ (for `constant,' $0 \rightarrow 0$ and $1 \rightarrow 0$ \emph{or} $0 \rightarrow 1$ and $1 \rightarrow 1$; for `balanced,'  $0 \rightarrow 0$ and $1 \rightarrow 1$ \emph{or} $0 \rightarrow 1$ and $1 \rightarrow 0$). Classically, a solution requires two queries to the oracle, for the input values 0 and 1, and a comparison of the outputs. 

Deutsch's algorithm begins by initializing 1-qubit input and output registers  to the state $\ket{0}\ket{0}$.  A Hadamard transformation ($\ket{0} \rightarrow \frac{1}{\sqrt{2}}(\ket{0} + \ket{1}$, $\ket{1} \rightarrow \frac{1}{\sqrt{2}}(\ket{0} - \ket{1})$) is applied to the input register (yielding a linear superposition of states corresponding to the two possible input values 0 and 1) followed by a unitary transformation $U_{f}$ applied to both registers that implements the Boolean function $f$:
\begin{eqnarray}
\ket{0}\ket{0} & \stackrel{H}{\rightarrow} & \frac{1}{\sqrt{2}}(\ket{0} + \ket{1})\ket{0} \\ 
& \stackrel{U_{f}}{\rightarrow} & \frac{1}{\sqrt{2}}(\ket{0}\ket{f(0)} + \ket{1}\ket{f(1)})
\end{eqnarray}

The final composite state of both registers is then one of  two orthogonal states, either (constant):
\begin{eqnarray}
\ket{c_{1}} & = & \frac{1}{\sqrt{2}}(\ket{0}\ket{0} + \ket{1}\ket{0}) \label{eqn:c1} \\
\ket{c_{2}} & = & \frac{1}{\sqrt{2}}(\ket{0}\ket{1} + \ket{1}\ket{1}) \label{eqn:c2}
\end{eqnarray}
or (balanced):
\begin{eqnarray}
\ket{b_{1}} & = & \frac{1}{\sqrt{2}}(\ket{0}\ket{0} + \ket{1}\ket{1}) \label{eqn:b1} \\
\ket{b_{2}} & = & \frac{1}{\sqrt{2}}(\ket{0}\ket{1} + \ket{1}\ket{0}) \label{eqn:b2}
\end{eqnarray}

The states $\ket{c_{1}}, \ket{c_{2}}$ and $\ket{b_{1}}, \ket{b_{2}}$  span two planes $P_{c}, P_{b}$  in 
$\hil{H}^{2}\otimes\hil{H}^{2}$, represented by the projection operators:
\begin{eqnarray}
P_{c} & = & P_{\ket{c_{1}}} \vee P_{\ket{c_{2}}} \\
P_{b} & = & P_{\ket{b_{1}}} \vee P_{\ket{b_{2}}}
\end{eqnarray}

Although the states $\ket{c_{1}}, \ket{c_{2}}$ are not orthogonal to the states $\ket{b_{1}}, \ket{b_{2}}$,  the  planes---which represent quantum disjunctions\footnote{Since $P_{\ket{c_{1}}}$ and $P_{\ket{c_{2}}}$ are orthogonal, $P_{c} = P_{\ket{c_{1}}} \vee P_{\ket{c_{2}}} = P_{\ket{c_{1}}} + P_{\ket{c_{2}}}$, where `$\vee$' represents quantum disjunction: the infimum or span (the smallest subspace containing the two component subspaces). Similarly for  $P_{b}$. Note that a classical disjunction is the infimum of the disjuncts in the Boolean algebra of classical logic.}---are orthogonal, except for an intersection, so their projection operators commute. The intersection is the line (ray) spanned by the vector:
 \begin{equation}
\frac{1}{2}(\ket{00} + \ket{01} + \ket{10} + \ket{11}) = \frac{1}{\sqrt{2}}(\ket{c_{1}} + \ket{c_{2}}) = \frac{1}{\sqrt{2}}(\ket{b_{1}} + \ket{b_{2}})
\end{equation}

In the prime basis spanned by the states $\ket{0'} = H\ket{0}, \ket{1'} = H\ket{1}$, the intersection is the state $\ket{0'}\ket{0'}$, the constant plane is spanned by:
\begin{eqnarray}
\ket{0'}\ket{0'} \\
\ket{0'}\ket{1'} & = & \frac{1}{\sqrt{2}}(\ket{c_{1}} - \ket{c_{2}})
\end{eqnarray} 
and the balanced plane is spanned by:
\begin{eqnarray}
\ket{0'}\ket{0'} \\
\ket{1'}\ket{1'} & = & \frac{1}{\sqrt{2}}(\ket{b_{1}} - \ket{b_{2}})
\end{eqnarray} 
i.e., 
\begin{eqnarray}
P_{c} & = & P_{\ket{0'}\ket{0'}} + P_{\ket{0'}\ket{1'}} \label{eqn:const} \\
P_{b} & = & P_{\ket{0'}\ket{0'}} + P_{\ket{1'}\ket{1'}} \label{eqn:bal}
\end{eqnarray}

To decide whether the function $f$ is constant or balanced we could measure the observable with eigenstates $\ket{0'0'}$, $\ket{0'1'}$, $\ket{1'0'}$, $\ket{1'1'}$ on the final state, which is in the 3-dimensional subspace orthogonal to the vector $\ket{1'0'}$, either in the constant plane or the balanced plane. If the state is in the constant plane,  we will either obtain the outcome $0'0'$ with probability 1/2 (since the final state is at an angle $\pi/4$ to $\ket{0'0'}$ and the probability is given by $\mbox{cos}^{2}(\frac{\pi}{4})$), in which case the computation is inconclusive, or the outcome $0'1'$ with probability 1/2. If the state is in the balanced plane, we will either obtain the outcome $0'0'$ with probability 1/2, in which case the computation is again inconclusive, or the outcome $1'1'$ with probability 1/2. So in either case, with probability 1/2, we can distinguish whether the function is constant or balanced  in one run of the algorithm by distinguishing between the constant and balanced planes, without evaluating the function at any of its inputs (i.e., without determining in the constant case whether $f$ maps 0 to 0 and 1 to 0, or whether $f$ maps  0 to 1 and 1 to 1, and similarly in the balanced case).\footnote{Equivalently, we could measure the output register. If the outcome is $0'$, the computation is inconclusive. If the outcome is $1'$, we measure the input register. The outcome $1'$ or $0'$ for the measurement on the input register then distinguishes whether the function is constant or balanced.} Recall that a classical algorithm requires two queries to the oracle, for the input values 0 and 1, and a comparison of the outputs. 

Note that, in evaluating the efficiency of the algorithm, we would have to take account of the fact that a computation is required to determine the prime basis, and we would have to count the number of computational steps required for this computation. We would also have to consider whether the measurement could be implemented efficiently. These sorts of issues might be trivial for the simple XOR algorithm, but in general will not be. To avoid this sort of problem, the number of relevant computational steps in a quantum algorithm is conventionally counted as the number of applications of unitary transformations and measurements required for the successful implementation of the algorithm, where the unitary transformations belong to a standard set of elementary unitary gates that form a universal set, and the measurements are in the computational basis. 

It is easy to see that a Hadamard transformation applied to the final states of both registers allows the constant  and balanced planes to be distinguished (again with probability 1/2) by a measurement in the computational basis. Since $H^{2} = I$, it follows that $\ket{0'0'} \stackrel{H}{\longrightarrow} \ket{00}$, etc., so a Hadamard transformation of the state amounts to dropping the primes in the representation (\ref{eqn:const}), (\ref{eqn:bal}) for the constant and balanced planes.

Since the quantum algorithm for Deutsch's XOR problem has an even probability of failing, the improvement in efficiency is only achieved if the algorithm succeeds, and even then is rather modest: one run of the quantum algorithm versus two runs of a classical algorithm. In a variation of the algorithm by Cleve \cite{Cleve98}, the output register is initialized to $\ket{1}$ instead of $\ket{0}$. 
Instead of the final state of the two registers ending up as one of two orthogonal states in the constant plane, or as one of two orthogonal states in the balanced plane, the final state now ends up as $\pm\ket{0'1'}$ in the constant plane, or as $\pm\ket{1'1'}$ in the balanced plane, and these states can be distinguished because they are orthogonal. So we can decide with certainty whether the function is constant or balanced after only one run of the algorithm. In fact, we can distinguish these two possibilities by simply measuring the input register in the prime basis, and since  a final Hadamard transformation on the state of the input register  takes $\ket{0'}$ to $\ket{0}$ and $\ket{1'}$ to 
$\ket{1}$), we can distinguish the two planes by measuring the input register in the computational basis. See \cite{Bub2007b} for details.

Deutsch's XOR problem can be generalized to the problem (`Deutsch's problem') of determining whether a Boolean function $f:\{0,1\}^{n} \rightarrow \{0,1\}$ is constant or whether it is balanced, where it is promised that the function is either constant or balanced.  `Balanced' here means that the function takes the values 0 and 1 an equal number of times, i.e., $2^{n-1}$ times each. Exploiting the Cleve variation of the XOR algorithm, the Deutsch-Jozsa algorithm \cite{DeutschJozsa92} determines whether $f$ is constant or balanced in one run and is exponentially faster than any classical algorithm. See \cite{Bub2007b}.

Deutsch's XOR problem and Deutsch's problem are elementary and perhaps not very interesting problems, but the quantum algorithms for their solution are structurally similar to Shor's factorization algorithm \cite{Shor94,Shor97}, perhaps the best-known quantum algorithm. Shor's algorithm, which achieves a remarkable exponential speed-up over any known classical algorithm, is essentially an algorithm for finding the period of a function. The algorithm exploits the fact that the two prime factors $p, q$ of a positive integer $N = pq$ can be found by determining the period of a function 
$f(x) = a^{x}\mbox{ mod $N$}$, for any $a < N$  which is coprime to $N$, i.e., has no common factors with $N$ (other than 1).  The period $r$ of $f(x)$ depends on $a$ and $N$. Once we know the period, we can factor $N$ if $r$ is even and $a^{r/2} \neq -1 \mbox{ mod $N$}$, which will be the case with probability greater than 1/2 if $a$ is chosen randomly. (If not, we choose another value of $a$.) The factors of $N$ are the greatest common factors of $a^{r/2} \pm 1$ and $N$, which can be found in polynomial time by the Euclidean algorithm. (For these number-theoretic results, see \cite[Appendix 4]{NielsenChuang}.) So the problem of factorizing a composite integer $N$ that is the product of two primes reduces to the problem of finding the period of a certain function $f: Z_{s} \rightarrow Z_{N}$, where $Z_{n}$ denotes the additive group of integers mod $n$.\footnote{Note that $f(x+r) = f(x)$ if $x+r \leq s$. The function $f$ is periodic if $r$ divides $s$ exactly, otherwise it is almost periodic. This adds a complication,  but does not change anything essential in the analysis.}

A particular period partitions the input values to the function into mutually exclusive and collectively exhaustive subsets: the subsets that are mapped onto the same output value by the periodic function. So the period of a function corresponds to a disjunctive property of the function: the disjunction over the different subsets of a particular partition. Distinguishing the period from alternative possible periods amounts to distinguishing the corresponding partition or disjunction from alternative possible partitions or disjunctions. A  classical algorithm that requires the evaluation of the function for a subset of input values to determine the partition involves an exponentially increasing number of steps as the size of the input increases. 

Shor's algorithm works as a period-finding algorithm by representing alternative partitions of the input values for a function, defined by alternative possible periods, as subspaces in a Hilbert space (i.e., quantum disjunctions), which are orthogonal except for overlaps (where the overlap region in some cases might include the entire subspace corresponding to a particular period). The subspace corresponding to a particular partition is spanned by orthogonal linear superpositions of states associated with the elements in  the (mutually exclusive and collectively exhaustive) subsets of the partition. The period-finding algorithm is designed to produce an entangled state in which such superpositions, representing orthogonal states of an input register, are correlated with distinct orthogonal states of an output register. The reduced state of the input register is then an equal-weight mixture of states spanning the subspace corresponding to the partition, where each state represents a subset in the partition as a linear superposition of the elements in the subset. Since the subspaces associated with different partitions are represented by commuting projection operators, a measurement of the state of the  input register in a certain basis can reveal the subspace containing the state, and hence the period associated with the partition, except when the measurement projects the state onto the overlap region (in which case the algorithm is run again until the outcome is definitive). This measurement basis is unitarily related to the computational basis by a known unitary transformation that can be implemented efficiently, so a measurement in the computational basis after this unitary information yields the same information. For details in terms of a specific  example, see \cite{Bub2007b}.

\section{Pseudo-telepathic Games}
\label{pseudo}

The notion of pseudo-telepathic games as a way of capturing the difference between \emph{classical correlations} among several parties, associated with common causes or `shared randomness,' and the \emph{quantum correlations} of entangled states shared by the parties, goes back to an unpublished idea of Allen Stairs\footnote{The idea is discussed in Stairs' PhD dissertation \cite{Stairs}.} exploited (with acknowledgement) by Heywood and Redhead \cite{Heywood}, and later by Greenberger, Horne, and Zeilinger \cite{GHZ} and Mermin \cite{Mermin1990}. The terminology was explicitly introduced in a paper by Brassard, Cleve, and Tapp \cite{Brassard1999}. A pseudo-telepathic game is a game between two or more separated players, who are not allowed to communicate after the game starts, that can be won if the players share entanglement, but cannot be won if the only shared resource consists of sequences of random numbers that label correlated responses (outputs) in the game and function, in effect, as common causes of the players' responses. The amount of communication required between players whose only resource is shared randomness to win the game is a measure of the difference between the quantum correlations of the players sharing entanglement and the classical correlations of the players sharing random numbers (or common causes, or local hidden variables).

A \emph{two-party game} between players Alice and Bob is defined as a sextuple $G = <X,Y,A,B,P,W>$, where $X$ and $Y$ are input sets, $A$ and $B$ are output sets, $P \subseteq X \times Y$ is a predicate on $X \times Y$ known as a promise, and $W \subseteq X \times Y \times A \times B$,  the winning condition, is a relation between inputs and outputs that is required to be satisfied by Alice and Bob, whenever the promise is fulfilled, in order to win the game (see \cite{Brassard2004}). Classical players are allowed to discuss strategy and exchange unlimited amounts of classical information, including random sequences, before the game starts. Quantum players are allowed to share unlimited entangled states. After the game starts, the players are separated and not allowed to communicate. In each round of the game, Alice and Bob are presented with inputs $x \in X$ and $y \in Y$, respectively. Alice and Bob win the round if either $(x,y) \notin P$, i.e., the promise is not fulfilled, or $(x,y,a,b) \in W$. Alice and Bob have a \emph{winning strategy} if it can be proved that they are certain to win the game for any possible round. Note that it is possible for Alice and Bob to win round after round without having a winning strategy, just as it is possible to toss `heads' round after round without having a biased coin. Of course, losing a single round shows that Alice and Bob do not have a winning strategy. A game is a \emph{pseudo-telepathic game} if there is no winning strategy for classical players (whose only allowed resource is shared randomness), but there is a winning strategy for quantum players (who are allowed to use entanglement as a resource). The term is suggested because, from a classical viewpoint, the success of the quantum strategy in such a case would seem to require `telepathic' communication between the players.

The following game (first proposed in \cite{Brassard1999}), called the Deutsch-Jozsa game because the quantum strategy is based on the Deutsch-Jozsa algorithm for Deutsch's problem, illustrates the idea. Alice and Bob are separately presented with bit strings $x$ and $y$, each of length $2^{m}$. They are required to output bits $a$ and $b$, respectively, each of length $m$, such that $a = b$ if and only if $x = y$. The promise is that $x = y$ or, if $x \neq y$, then the sequences $x$ and $y$ differ in exactly half the positions. 

Consider the problem with $m = 1$. Alice is presented with a string $x = x_{0}x_{1}$, where $x_{0} \in \{0,1\}$ and  $x_{1} \in \{0,1\}$. Simlarly, Bob is presented with a string $y = y_{0}y_{1}$, where $y_{0} \in \{0,1\}$ and  $y_{1} \in \{0,1\}$. Alice and Bob are required to output bits $a \in \{0,1\}$ and $b \in \{0,1\}$, respectively, such that $a = b$ if $x = y$, and $a \neq b$ if $x \neq y$.

It is easy to see that, in this case ($m = 1$),  there is a classical strategy for winning the game, even without exploiting shared randomness, because  the promise entails that when $x \neq y$:
\begin{equation}
x = 00 \mbox{ or }  11 \mbox{ if and only if } y = 01 \mbox{ or }  10 \label{eqn:promise}
\end{equation}
For example, the input $x =00, y = 11$ conflicts with the promise.
So if Alice and Bob output:
\begin{eqnarray}
a & = & x_{0} \oplus x_{1}  \\
b & = & y_{0} \oplus y_{1}
\end{eqnarray}
where $\oplus$ is Boolean addition (addition modulo 2), then:
\begin{equation}
a = b \mbox{ if and only if } x = y
\end{equation}

This strategy works because, given the premise, the question of whether $x = y$ or $x \neq y$ (which requires comparing two \emph{pairs} of bits) reduces to the question of whether the parity of $x$ is the same as the parity of $y$ (which, given the two parities, requires comparing two bits). Note that (\ref{eqn:promise})---and (different) parity---can be expressed as a disjunctive property of the input strings. The strategy employed by Alice and Bob succeeds in winning the game, and is guaranteed to succeed, even though the strategy provides no information to Bob about Alice's input $x$, or to Alice about Bob's input $y$. We could think of the strategy as a two-party classical computational algorithm that reduces the problem of deciding whether two 2-bit strings held separately by the two parties are the same or different, given the promise, to the problem of deciding whether two 1-bit strings are the same or different, \emph{without providing either party any information about the identity of the individual bits held by the other party}.

Consider, now, the corresponding quantum strategy. Alice and Bob start with the two-qubit shared entangled state:\footnote{That is, before the game starts they prepare and share many labeled copies of the entangled state $\ket{\Psi}_{i}$, so that they are both able to use the $i$'th copy for the $i$'th round of the game.}
\begin{equation}
\ket{\Psi} = \frac{1}{\sqrt{2}}(\ket{0}\ket{0} + \ket{1}\ket{1})
\end{equation}
Alice applies the unitary transformation:
\begin{eqnarray}
\ket{0} & \stackrel{U_{A}}{\rightarrow} & (-1)^{x_{0}}\ket{0} \\
\ket{1} & \stackrel{U_{A}}{\rightarrow} & (-1)^{x_{1}}\ket{1}
\end{eqnarray}
to her qubit, followed by a Hadamard transformation ($\ket{0} \stackrel{H}{\rightarrow} 1/\sqrt{2}(\ket{0} + \ket{1}), \ket{1} \stackrel{H}{\rightarrow} 1/\sqrt{2}(\ket{0} - \ket{1})$), and similarly Bob applies the unitary transformation:
\begin{eqnarray}
\ket{0} & \stackrel{U_{B}}{\rightarrow} & (-1)^{y_{0}}\ket{0} \\
\ket{1} & \stackrel{U_{B}}{\rightarrow} & (-1)^{y_{1}}\ket{1}
\end{eqnarray}
to his qubit followed by a Hadamard transformation. Alice and Bob then measure the transformed state in the computational basis. The respective measurement outcomes provide the outputs $a$ and $b$ for the strategy. It is easy to see that the outputs satisfy the condition for winning the game:
\begin{equation}
a = b \mbox{ iff } x = y
\end{equation}

For example, suppose $x = 00, y = 01$, so $x \neq y$. The shared entangled state undergoes the transformations:
\begin{eqnarray}
\frac{1}{\sqrt{2}}(\ket{0}\ket{0} + \ket{1}\ket{1}) & \stackrel{U_{A}U_{B}}{\rightarrow} & \frac{1}{\sqrt{2}}(\ket{0}\ket{0} - \ket{1}\ket{1}) \\
& \stackrel{H_{A}H_{B}}{\rightarrow} & \frac{1}{\sqrt{2}}(\ket{0}\ket{1} + \ket{1}\ket{0})
\end{eqnarray}
A measurement by Alice and Bob on this state produces the outcomes $a = 0, b = 1$ or $a = 1, b = 0$, i.e., $a \neq b$.

As with the classical algorithm, the quantum algorithm succeeds without providing either party any information about the identity of the individual bits held by the other party. 

Notice that the quantum algorithm produces outputs $a$ and $b$ that are randomly related to the parity of the corresponding inputs (i.e., when the parity of Alice's input string is 0, she outputs 0 or 1 with equal probability, and similarly when the parity is 1; the same goes for Bob). So we should really add this constraint to the winning condition in the case $m=1$. In this case, there is still a classical winning strategy for the game, but the players would now need to exploit a shared random bit string $\lambda_{i} \in \{0,1\}$, where the bit $\lambda_{i}$ is used in obtaining the output for each round $i$ of the game. A winning classical strategy would be for Alice and Bob to output, for each round $i$:
\begin{eqnarray}
a & = & \lambda_{i} \oplus x_{0} \oplus x_{1}  \\
b & = & \lambda_{i} \oplus y_{0} \oplus y_{1}
\end{eqnarray}
Here  $\lambda_{i}$ is  a hidden variable labeling a common cause of the correlated outputs in each round of the game.

The Deutsch-Jozsa game can be solved by a classical strategy for the case $m = 1$, $m = 2$, and $m = 3$  (see \cite{Galliard2002}), but there exists no classical strategy for $m = 4$ (see \cite{Galliard2003}). For $m$ sufficiently large, classical players cannot win the Deutsch-Jozsa game without communicating at least $c2^{m}$ bits, for some appropriate constant $c > 0$ (see \cite{Buhrman1998}). The quantum strategy succeeds for all values of $m$, reducing the problem of deciding whether two privately held $2^{m}$-bit strings are the same or different, given the promise, to the problem of deciding whether two $m$-bit strings are the same or different, without transferring any information between the two parties. 

In effect, the quantum strategy provides a protocol for a secure two-party computation. For example, suppose Alice and Bob are adversaries who want to decide whether their data bases are the same or different, but they would like to keep as much of the data private as they can in case it turns out that the data bases are different. The protocol allows a determination for data bases of length $n = 2^{m}$ on the basis of an exponentially smaller amount of just $m$ bits of shared information (under the constraint provided by the promise).

For the general case, $n = 2^{m}$, Alice and Bob start with the shared entangled state in an $n \times n$-dimensional Hilbert space:
\begin{equation}
\ket{\Psi} = \frac{1}{\sqrt{n}}\sum_{j = 0}^{n-1}\ket{j}\ket{j}
\end{equation}
Taking Alice's Hilbert space as a tensor product of $m$ qubits (i.e., $2^{m}$-dimensional), and similarly Bob's, the entangled state $\ket{\Psi}$ can be expressed as a linear superposition of $m \times m$ product qubit states for Alice and Bob by reading the index $j$ as a binary integer. So, for example, for $n = 2^{2}$, i.e., $m = 2$, where the (Alice+Bob)-Hilbert space $\hil{H}^{16}$ is $4 \times 4$-dimensional, we have:
\begin{eqnarray}
\ket{\Psi} & = & \frac{1}{2^{2}}(\ket{0}\ket{0} + \ket{1}\ket{1} + \ket{2}\ket{2} + \ket{3}\ket{3}) \\
& = & \frac{1}{2^{2}} (\ket{00}\ket{00} + \ket{01}\ket{01} + \ket{10}\ket{10} + \ket{11}\ket{11})
\end{eqnarray}
The effect of the unitary transformations of the quantum algorithm is to leave $\ket{\Psi} \in \hil{H}^{16}$ unchanged if $x = y$ (in which case the measurements yield $a = b$), or,  if $x \neq y$, the effect is to move $\ket{\Psi}$ to an 8-dimensional subspace of $\hil{H}^{16}$ which is orthogonal to the 4-dimensional subspace containing $\ket{\Psi}$  spanned by $\ket{00}\ket{00}, \ket{01}\ket{01}, \ket{10}\ket{10}, \ket{11}\ket{11}$ and orthogonal to the 4-dimensional subspace forbidden by the promise, spanned by $\ket{00}\ket{11}, \ket{11}\ket{00}, \ket{10}\ket{01}, \ket{11}\ket{01}$.

\section{Conclusion}

A quantum algorithm succeeds not because the superposition principle allows `the computation of all values of a function at once' via `quantum parallelism,' but rather because the structure of a quantum state space allows new sorts of correlations associated with entanglement, with new possibilities for information-processing transformations between correlations,  that are not possible in a classical state space.

Classical states, pure and mixed, form a convex set that has the very specific structure of a \emph{simplex}. For a finite set of extremal points or pure states in an affine space---e.g., a set of pure states that can be specified by finite bit strings---the smallest convex set that contains the pure states (the `convex hull' of the pure states) is called a convex polytope. A $p$-simplex is the convex polytope of $p+1$ pure states not confined to any $(p-1)$-dimensional subspace (e.g., a 2-simplex is a triangle on a plane, with the vertices representing pure states and all other points, in the interior and on the edges, representing mixed states). A simplex has the very special property that any point in the simplex can be expressed as a mixture (probability distribution, convex combination) of pure states in one and only one way \cite{Bengtsson2006}. For example, if there are only two pure states, denoted by $\stackrel{\rightarrow}{0}$ and $\stackrel{\rightarrow}{1}$, the simplex is the line joining the two pure states, and any point $\stackrel{\rightarrow}{p}$ on the line can be represented in one and only one way as a mixture of the pure states $\stackrel{\rightarrow}{0}$ and $\stackrel{\rightarrow}{1}$: 
\begin{equation}
\stackrel{\rightarrow}{p} = \lambda_{0}\stackrel{\rightarrow}{0} + \lambda_{1}\stackrel{\rightarrow}{1},
(\lambda_{0},  \lambda_{1},   \geq 0,  \lambda_{0}  + \lambda_{1} = 1)
 \end{equation}
 This is a property of a simplex, i.e., a feature of the space of (pure or mixed) classical states or classical probability measures, and no other convex set has this property. In particular, it is not a property of the space of quantum states or quantum probability measures, which does not have the structure of a simplex. 

The Bell inequalities are a characteristic feature of correlations associated with a simplex structure \cite{Pitowsky1989b}. The Clauser-Horne-Shimony-Holt  version of the Bell inequalities for a bipartite classical system requires a certain correlation function to take values less than or equal to 2 \cite{CHSH}. For an entangled bipartite quantum system,  this correlation can take a maximum value of $2\sqrt{2}$, the Tsirelson bound \cite{Tsirelson1980}. What this means, essentially, is that new sorts of correlations are possible for quantum systems that are not possible in a classical, simplex structure.

In a quantum computation, the information-processing of a quantum algorithm transforms correlations represented by the quantum state of the input and output registers (in a way that would be impossible in a classical simplex structure)  to determine a disjunctive property of a function, without the information-processing requiring a determination of any of the individual disjuncts in the disjunction. This is analogous to the $m = 1$ Deutsch-Jozsa game which has a classical solution, where Alice and Bob can output the same 1-bit string if and only if the 2-bit input strings are the same, without either party coming to know the bits in the other party's input string. The classical solution for $m = 1$ depends on the possibility of local information-processing of the input strings to produce a correlation in the winning condition. For $m = 4$, the game becomes a pseudo-telepathic game, with a quantum strategy but no classical strategy. The quantum strategy exploits the possibility of correlations in the winning condition $W \subseteq X \times Y \times A \times B$ arising from local operations on the quantum state $\ket{\Psi}$ that are not possible in a classical simplex theory. 

The question of the efficiency of a quantum algorithm for a particular problem, relative to any possible classical algorithm, is a further issue that requires consideration of the relevant computational steps in the quantum algorithm compared to the computational steps in a classical algorithm. The purpose of the preceding analysis was to point out the inadequacy of the `quantum parallelism' view of what is distinctive about a quantum computation, and to propose a very different view.

\bibliographystyle{plain}
\bibliography{ComputationPSA}

\end{document}